\documentclass[11pt]{article}
\usepackage{moriond,epsfig}
\usepackage{xspace}
\usepackage{caption}

\def\MyPhoto{\psfig{figure=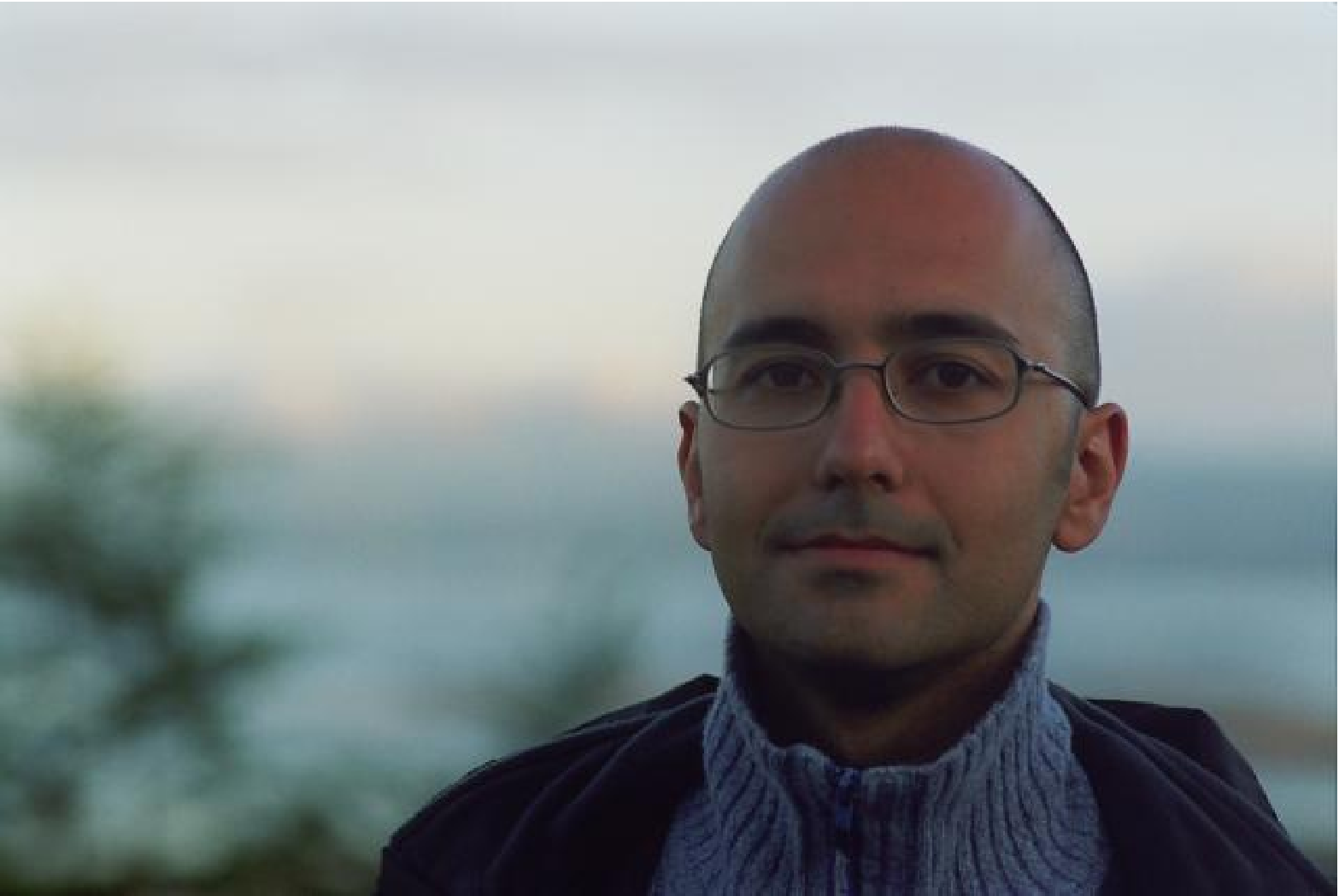,width=35mm}}

\def\Journal#1#2#3#4{{#1} {\bf #2}, #3 (#4)}

\newcommand{\nn}{\ensuremath{\rm NN}\xspace}

\newcommand{\ke}{\ensuremath{K \to e \nu}\xspace }
\newcommand{\km}{\ensuremath{K \to \mu \nu}\xspace}

\newcommand{\ked}{\ensuremath{K_{e2}}\xspace}
\newcommand{\kedg}{\ensuremath{K_{e2\gamma}}\xspace}
\newcommand{\ket}{\ensuremath{K_{e3}}\xspace}
\newcommand{\kmt}{\ensuremath{K_{\mu3}}\xspace}

\newcommand{\kmd}{\ensuremath{K_{\mu2}}\xspace}

\begin{document}
\vspace*{4cm}
\title{PRECISION TEST OF THE SM WITH Kl2 AND Kl3 DECAYS AT KLOE}

\author{ T. SPADARO }

\address{Laboratori Nazionali di Frascati dell'INFN, Via E. Fermi, 40,\\
00044 Frascati (Roma), Italy}

\maketitle\abstracts{
Kaon decay studies seeking new-physics (NP)
effects in leptonic ($K_{l2}$) or semileptonic ($K_{l3}$) decays are discussed. 
A unitarity test of the first row of the
CKM mixing matrix is obtained from the KLOE precision measurements 
of Kl3 widths for $K^\pm$, $K_L$, and (unique to KLOE) $K_S$, complemented with the absolute branching ratio for the $K_{\mu2}$ decay.
KLOE results lead to constraints for NP models and can probe
possible charged Higgs exchange contribution in SM extensions with 
two Higgs doublets.
The main focus in the present document is set on a new measurement of $R_K=\Gamma(K_{e2})/\Gamma(K_{\mu2})$ with an accuracy at the \% level,
aiming at finding evidence of deviations from the SM prediction induced by
lepton-flavor violation NP effects.}

\section{Introduction}
%
%
%
New precise measurements of $K\to l\nu_{l}(\gamma)$ ($K_{l2}$) and $K\to \pi l\nu_{l}(\gamma)$ ($K_{l3}$) decays can possibly shed 
light on new physics (NP). The first indication of the need of improving the present knowledge in this field
was given by the 2004 version of the PDG: a deviation from unitarity of the CKM matrix was observed in the first row, amounting
to more than two standard deviations~\cite{PDG+04},
\begin{equation}
\Delta = 1-V_{ud}^2-V_{us}^2-V_{ub}^2 = 0.0043(16)_{Vud}(11)_{Vus}.
\end{equation}
This called for new precise determinations of the $V_{us}$ parameter of the CKM matrix, traditionally
extracted from $K_{l3}$ decays using the following expression:
\begin{equation}\nonumber
\Gamma^i(K_{e3(\gamma),\,\mu3(\gamma)}) = |V_{us}|^2\frac{C_i^2 G_F^2 M^5}{128\pi^3} S_{\rm
EW}\:  |f^{K^0}_+(0)|^2 I^{i}_{e3,\,\mu3}\: 
  (1+\delta_{e3,\,\mu3}^{i}),
\end{equation}
where $i$ indexes $K^0\to\pi^-$ and $K^+\to\pi^0$ transitions for which $C_i^2 =1$ and 1/2, respectively, $G_F$ is the Fermi
constant, $M$ is the appropriate kaon mass, and $S_{\rm EW}$ is a
universal short-distance electroweak correction~\cite{as}. 
The $\delta^i$ term accounts for long-distance radiative corrections depending on the 
meson charges and lepton masses and, for $K^\pm$,
for isospin-breaking effects. These corrections are presently known at the few-per-mil level~\cite{aa}.
The $f^{K^0}_+(0)$ form factor parametrizes the vector-current transition 
$K^0\to\pi^-$ at zero momentum transfer $t$, while the dependence of
vector and scalar form factors on $t$ enter into the determination of the integrals $I_{e3,\,\mu3}$
of the Dalitz-plot density over the physical region.

After four years of analysis of KLOE data, we present the most
comprehensive set of results from a single experiment, including 
BR's for $K_{e3}$ and $K_{\mu3}$ decays for $K_{L}$~\cite{KLOE+06:KLBR} and $K^\pm$~\cite{KLOE+08:KPBR}, 
and the BR for $K_S\to\pi e\nu$~\cite{KLOE+06:KSe3,KLOE+06:KSratio} (unique to KLOE); 
form factor slopes from analysis of $K_{Le3}$~\cite{KLOE+06:KLe3FF} and $K_{L\mu3}$~\cite{KLOE+08:KLm3FF}; 
lifetime measurements for $K_L$~\cite{KLOE+05:KLlife} and $K^\pm$~\cite{KLOE+08:KPlife}; 
the $K^0$ mass~\cite{KLOE+07:K0mass}. Using the $K_S$ lifetime from PDG~\cite{PDG+08} 
as the only input other than KLOE measurements, we obtain five results for the product $f_+(0)|V_{us}|$~\cite{KLOE+08:Vus},
as shown in table~\ref{tab:f0vus}. 
The average of these has been obtained taking all correlations into account and it is $f_+(0)\times|V_{us}|=0.2157(6)$. As a comparison, using data from KLOE, KTeV, NA48, and ISTRA+ 
experiments, the world average~\cite{Flavia+08} is $0.2166(5)$.
From the KLOE result and using $f_+(0)=0.9644(49)$ from the UKQCD/RBC collaboration~\cite{RBCUKQCD+07:f0}, we obtain 
\begin{equation}
\label{eq:kloevus}
|V_{us}|=0.2237(13).
\end{equation} 
Using the world average~\cite{Towner-Hardy07} $V_{ud}=0.97418(26)$ 
from $0^+\to0^+$ nuclear $\beta$ decays, 
CKM unitarity can be seen to be satisfied: $\Delta = 9(8)\times10^{-4}$.

KLOE has provided the most precise determination of the $K_{\mu2}$
BR~\cite{KLOE+06:BRKM2}, which can be linked to the ratio $V_{us}/V_{ud}$ via the following 
relation~\cite{Marciano+04:VusVud}:
\[
\frac{\Gamma(K\to\mu\nu)}{\Gamma(\pi\to\mu\nu)} = 
\frac{m_K\left(1-m_\mu^2/m_K^2\right)^2}{m_\pi\left(1-m_\mu^2/m_\pi^2\right)^2}
\left|\frac{V_{us}}{V_{ud}}\right|^2
\frac{f_K^2}{f_\pi^2}
C.
\]
The theoretical inputs are the form-factor ratio $f_K/f_\pi$ and the radiative corrections described by the factor $C$. We use $f_K/f_\pi=1.189(7)$ 
from lattice calculations~\cite{HPUKQCD+07:fkfp} and
$C=0.9930(35)$~\cite{Marciano+04:VusVud}, thus obtaining 
\begin{equation}
\label{eq:kloevusvud}
|V_{us}/V_{ud}|=0.2326(15).
\end{equation}
From the KLOE results of Eqs.~\ref{eq:kloevus} and~\ref{eq:kloevusvud}, and from the world-average
value of $V_{ud}$, a combined fit to $V_{us}$ and $V_{ud}$ has been done. 
The result is shown in left panel of Fig.~\ref{fig:vusvudfit}: 
the fit $\chi^2$ is 2.34 for one degree of freedom (13\% probability)
and the results are:
$|V_{us}|=0.2249(10)$ and $|V_{ud}|=0.97417(26)$,
with a correlation of 3\%.
From these, not only can we now state that the CKM unitary holds to within $10^{-3}$,
$\Delta = 0.0004\pm0.0005_{Vud}\pm0.0004_{Vus}$,
but we can obtain severe constraints for many NP models. 

\subsection{Unitarity and coupling-universality tests}

In the SM, unitarity of the weak couplings and gauge universality dictate:
\begin{equation}
G^2_\mathrm{F}\left(|V_{ud}|^2+|V_{us}|^2\right)=G^2_\mu \mbox{ }\left(V_{ub}^2\mbox{ negligible}\right),
\end{equation}
where $G^2_\mu$ is the decay constant obtained from the measurement of the $\mu$ 
lifetime~\cite{MuLan}. The above measurement of $V_{us}^2$ from KLOE inputs provides 
relevant tests for possible breaking 
of the CKM unitarity ($\Delta\neq 0$) and/or of the coupling universality ($G_F\neq G_\mu$). 
This can happen in some NP scenarios, some example of which we discuss below.

NP might lead to exotic and still unobserved $\mu$ decays contributing to the $\mu$ lifetime.
The resulting total BR for $\mu$ exotic modes equals the unitarity violation $\Delta$.
Some of these modes, such as $\mu^+\to e^+\overline{\nu}_{e}\nu_\mu$, are at present constrained to be less than $\sim1\%$,
so that information from unitarity improves on that from direct searches by more than 
a factor of 10~\cite{PDG+08:vusvud,Marciano+KAON07}.

The existence of additional heavy Z bosons would influence unitarity at the loop level entering in muon and charged current 
semileptonic decays differently~\cite{Marciano-Sirlin+87}: $\Delta = -0.01\lambda\ln[r^2_Z/(r^2_Z-1)]$, where
$r_Z=m_{Z^\prime}/m_W$ and 
$\lambda$ is a model-dependent constant of order 1. 
In the case of $SO(10)$ grand unification, $\lambda\sim1.9$ and a 
unitarity test from KLOE results yields $M_{Z^\prime}>750$~GeV at 95\% of CL. In non-universal gauge interaction models, a tree-level
contribution from $Z^\prime$ bosons appears, so that the unitarity test is sensitive to even larger masses~\cite{Lee+07}.

In supersymmetric extensions of the SM (SUSY), loops affect muon and semileptonic decays differently. Unitarity can constrain 
SUSY up to mass scales of the order of 0.5~TeV, depending on the extent of cancellation between squark and 
slepton effects~\cite{SUSY-cit}.

Measurements of $K_{l2}$ widths can be linked to new physics effects, too. 
The ratio of $K_{\mu2}$ to $\pi_{\mu2}$ decay widths might accept
NP contributions from charged Higgs exchange~\cite{hou,isidoriparadisi} 
in supersymmetric 
extensions of the SM with two Higgs doublets. 
In this scenario, the ratio $V_{us}/V_{ud}$ extracted from $K_{\mu2}$, $\pi_{\mu2}$ should differ
from that extracted from $K_{l3}$ and superallowed Fermi transitions (``$0^+$''):
\[
\left|\frac
{V_{us}(K_{l2})V_{ud}(0^+)}{V_{us}(K_{l3})V_{ud}(\pi_{l2})}
\right|=\left|1-\frac{m^2_{K}(m_s-m_d)\tan^2\beta}{M^2_{H}m_s(1+\epsilon_0\tan\beta)}
\right|,
\]
where $\tan\beta$ is the ratio of up- and down-Higgs vacuum
expectation values, $M_{H}$ is the charged Higgs mass, and
$\epsilon_0\sim0.01$~\cite{isidoriretico}.  The KLOE result of
Eq.~\ref{eq:kloevusvud} can be translated into an exclusion plot in
the plane $\tan\beta$ vs $M_H$ (see right panel of
Fig.~\ref{fig:vusvudfit}), showing that this analysis is complementary
to and competitive with that~\cite{isidoriparadisi} using the average
$\mathrm{BR}(B\to\tau\nu)=1.73(35)\times10^{-4}$ of Babar and Belle
measurements~\cite{bellebabar}.

\parbox[t]{0.3\textwidth}{
  \begin{tabular}{|l|l|l|}\hline
    Mode         & $f_+\times|V_{us}|$ & Error,\%    \\ \hline 
    $K_{Le3}$     & 0.2155(7)       &  0.3         \\
    $K_{L\mu3}$   & 0.2167(9)       &  0.4         \\
    $K_{Se3}$     & 0.2153(14)      &  0.7         \\
    $K^\pm_{e3}$  & 0.2152(13)      &  0.6         \\
    $K^\pm_{\mu3}$& 0.2132(15)      &  0.7          \\ \hline
  \end{tabular}
  \captionof{table}{\label{tab:f0vus}Five determinations of
    $f_+\times|V_{us}|$ using the $K_S$ lifetime (from PDG) as the
    only input other than KLOE measurements.}  }
\hfill
\parbox[t]{0.6\textwidth}{
  \begin{tabular}[c]{cc}
    \includegraphics[totalheight=3.6cm]{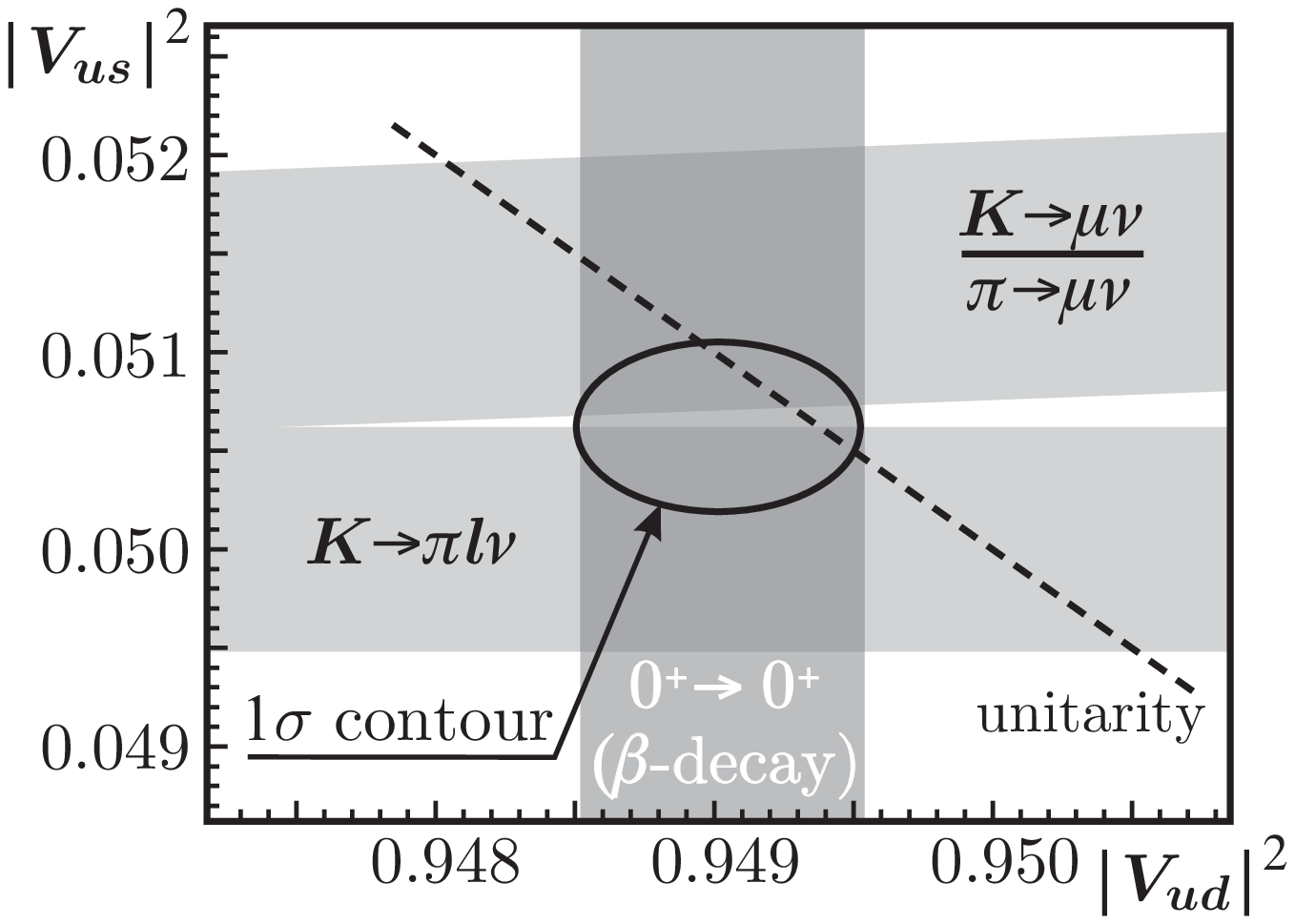} &
    \includegraphics[totalheight=3.6cm]{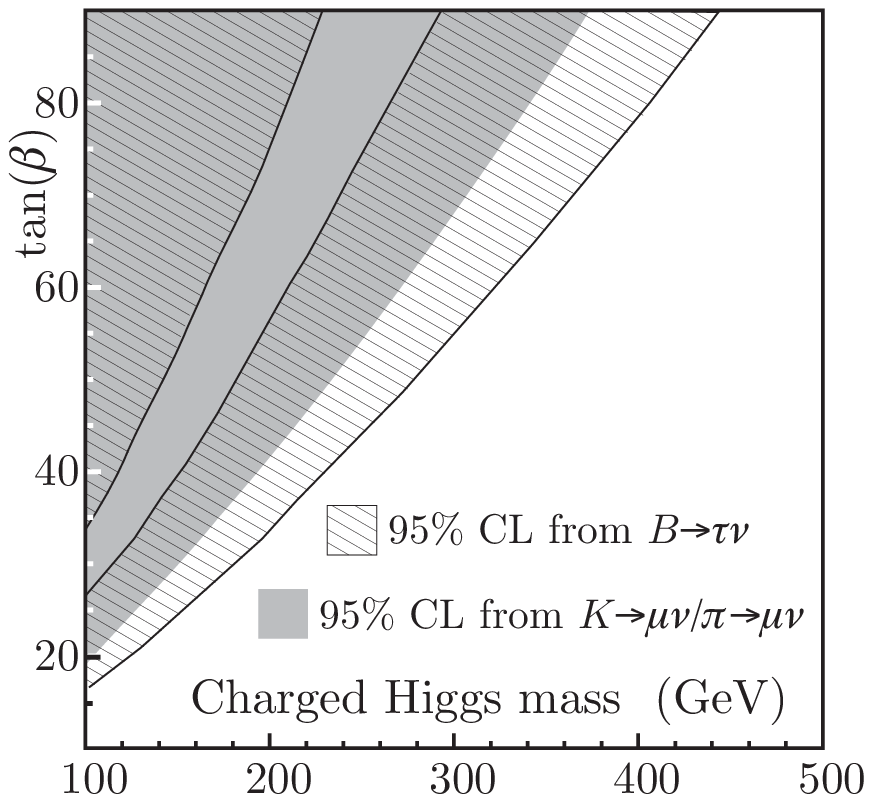} \\
  \end{tabular}
  \captionof{figure}{\label{fig:vusvudfit}Left: The 1-$\sigma$ fit
    result to $V_{ud}$ and $V_{us}$ is shown by the solid line
    ellipse, in agreement with the unitarity bound shown by the dashed
    line.  Right: Excluded regions from analysis of decays
    $K\to\mu\nu$ (filled area) and $B\to\tau\nu$ (hatched area).}  }


\subsection{Test of lepton-flavor violation}

A significant effort has been devoted along the years to isolate signals from
lepton flavor violating (LFV) transitions, which are forbidden or ultra-rare
in the Standard Model (SM). The sensitivity to decays such as $\mu\to e\gamma$,
$\mu\to eee$, $K_L\to \mu e(+\pi^0\mbox{'s})$, and others roughly improved by
two orders of magnitude for each decade~\cite{landsbergf04}. 
No signal has been observed, thus ruling out
SM extensions with LFV amplitudes with mediator masses below $\sim100$~TeV.

These results allowed the focus to be put on the detection of NP-LFV effects 
in loop amplitudes, by studying specific processes suppressed in the SM.
In this field, a strong interest for a new measurement of the ratio
$R_K=\Gamma(\ke)/\Gamma(\km)$
has recently arisen, triggered by the work of Ref.~\cite{masiero}. 
The SM prediction of $R_K$ benefits from
cancellation of hadronic uncertainties to a large extent and therefore
can be calculated with high precision. Including radiative
corrections, the total uncertainty is less than 0.5 per
mil~\cite{ciriglianorosell07}:
\begin{equation}
\label{eq:rkSM}
 R_K = (2.477\pm0.001)\times 10^{-5}.
\end{equation}
Since the electronic channel is
helicity-suppressed by the $V-A$ structure of the charged weak
current, $R_K$ can receive contributions from physics beyond the SM,
for example from multi-Higgs effects inducing an effective
pseudoscalar interaction.  It has been shown in Ref.~\cite{masiero}
that deviations from the SM of up to few percent on $R_K$ are quite
possible in minimal supersymmetric extensions of the SM and in
particular should be dominated by lepton-flavor violating
contributions with tauonic neutrinos emitted in the electron channel:
\begin{equation}
\label{eq:rkmssm}
R_K=R_K^{\mathrm{SM}}\times\left[1+ \left(\frac{m_K^4}{m_H^4}\right)
  \left(\frac{m^2_\tau}{m^2_e}\right)
  \left|\Delta_R^{31}\right|^2\tan^6\beta\right],
\end{equation}
where $M_H$ is the charged-Higgs mass, $\Delta_R^{31}$ is the effective $e$-$\tau$
coupling constant depending on MSSM parameters, and $\tan\beta$ is the
ratio of the two vacuum expectation values.  Note that the
pseudoscalar constant $f_K$ cancels in $R_K^{\mathrm{SM}}$.

In order to compare with the SM prediction at this level of accuracy,
one has to treat carefully the effect of radiative corrections, which
contribute to nearly half the \kedg width.  In particular, the
SM prediction of Eq.~\ref{eq:rkmssm} is made considering all
photons emitted by the process of internal bremsstrahlung (IB) while
ignoring any contribution from structure-dependent direct emission
(DE). Of course both processes contribute, so in the analysis
DE is considered as a background which can be distinguished from the IB
width by means of a different photon energy spectrum.

Two experiments are participating in the challenge to push the error on $R_K$ from the present
6\% down to less than 1\%. In 2007, KLOE and NA48/2 announced preliminary 
results~\cite{KLOEeNA48+07:ke2} with errors ranging from 2\% to 3\%.
Moreover, the new NA62 collaboration collected more than 100\,000 $K_{e2}$ events in a 
dedicated run of the NA48 detector, aiming at reaching an accuracy of few per mil on $R_K$~\cite{Winhart-moriond+09}.

\section{Measuring $R_K$ at KLOE}
DA$\Phi$NE, the Frascati $\phi$ factory, is an $e^{+}e^{-}$ collider
working at $\sqrt{s}\sim m_{\phi} \sim 1.02$~GeV. $\phi$ mesons are produced,
essentially at rest, with a visible cross section of $\sim$~3.1~$\mu$b
and decay into $K^+K^-$ pairs with a BR of $\sim 49$\%.

Kaons get a momentum of $\sim$~100~MeV/$c$ which translates into a low speed, $\beta_{K} \sim$ 0.2.
$K^+$ and $K^-$ decay with a mean length of $\lambda_\pm\sim $~90~cm and can be 
distinguished from their decays in flight to one of the two-body final states 
$\mu\nu$ or $\pi\pi^0$.

The kaon pairs from $\phi$ decay are produced in a pure $J^{PC}=1^{--}$ quantum state, so that 
observation of a $K^+$ in an event signals, or tags, the presence of a $K^-$
and vice versa; highly pure and nearly monochromatic $K^\pm$
beams can thus be obtained and exploited to achieve high precision in the measurement of 
absolute BR's.

The analysis of kaon decays is performed with the KLOE detector, consisting essentially of a drift chamber, DCH, surrounded by an
electromagnetic calorimeter, EMC. A superconducting coil provides a 0.52~T magnetic field.
The DCH~\cite{nimdch} is a cylinder of 4~m in diameter
and 3.3~m in length, which constitutes a fiducial volume 
for $K^\pm$ decays extending for $\sim1\lambda_\pm$.
The momentum resolution for tracks 
at large polar angle is $\sigma_{p}/p \leq 0.4$\%. 
The c.m.\ momenta reconstructed from identification of 1-prong $K^\pm\to\mu\nu,\pi\pi^0$ decay vertices in the DC 
peak around the expected values with a resolution of 1--1.5~MeV, thus allowing clean and efficient $K^\mp$ tagging. 

The EMC is a lead/scintillating-fiber sampling calorimeter~\cite{nimcalo}
consisting of a barrel and two endcaps, with good
energy resolution, $\sigma_{E}/E \sim 5.7\%/\sqrt{\rm{E(GeV)}}$, and excellent 
time resolution, $\sigma_{T} =$~54~ps$/\sqrt{\rm{E(GeV)}} \oplus 50$ ps. 
%

In early 2006, the KLOE experiment completed data taking, having collected
$\sim2.5$~fb$^{-1}$ of integrated luminosity at the $\phi$ peak,
corresponding to $\sim$3.6 billion $K^+K^-$ pairs.
Using the present KLOE
dataset, a measurement of $R_K$ with an accuracy of about 1~\% has been performed.

Given the $K^\pm$ decay length of $\sim$90~cm, the selection of
one-prong $K^{\pm}$ decays in the DC required to tag $K^{\mp}$ has an
efficiency smaller than 50\%. In order to keep the statistical
uncertainty on the number of \ke counts below 1\%,
a ``direct search'' for \ke and \km decays is performed, without
tagging. Since the wanted observable is a ratio of BR's for two channels with
similar topology and kinematics, one expects to benefit from some
cancellation of the uncertainties on tracking, vertexing, and
kinematic identification efficiencies.  Small deviations in the
efficiency due to the different masses of $e$'s and $\mu$'s will be
evaluated using MC.

Selection starts by requiring a kaon track decaying in a DC fiducial
volume (FV) with laboratory momentum between 70 and 130~MeV, and a
secondary track of relatively high momentum (above 180~MeV).
The FV is defined as a cylinder parallel to the beam axis with length
of 80~cm, and inner and outer radii of 40 and 150~cm, respectively.
Quality cuts are applied to ensure good track fits.

\begin{figure}[htbp]
  \centering
  \hspace*{0.in}\begin{minipage}[c]{0.4\textwidth}
    \begin{tabular}[c]{cc}
\includegraphics[totalheight=3.5cm]{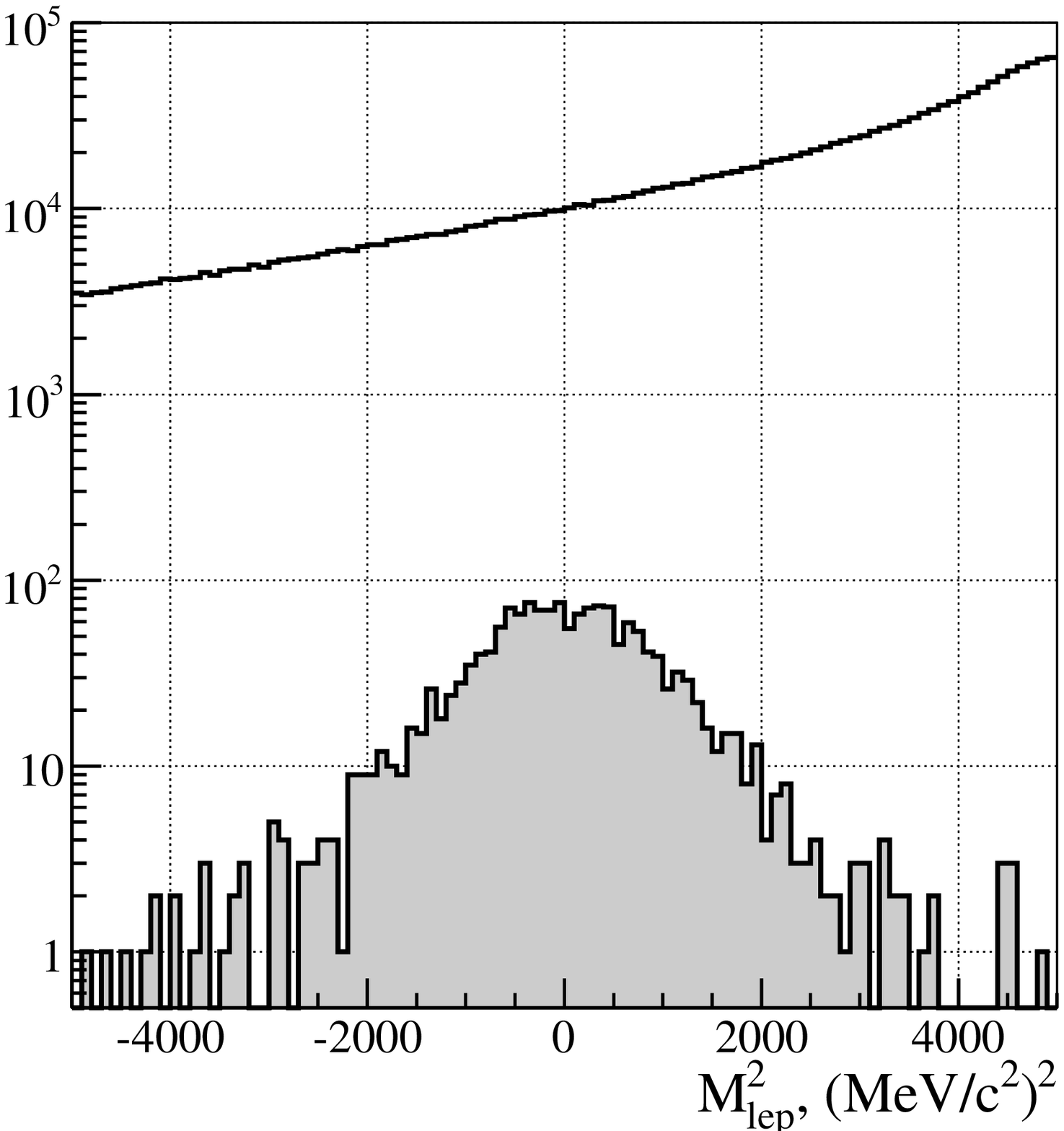}&
\includegraphics[totalheight=3.5cm]{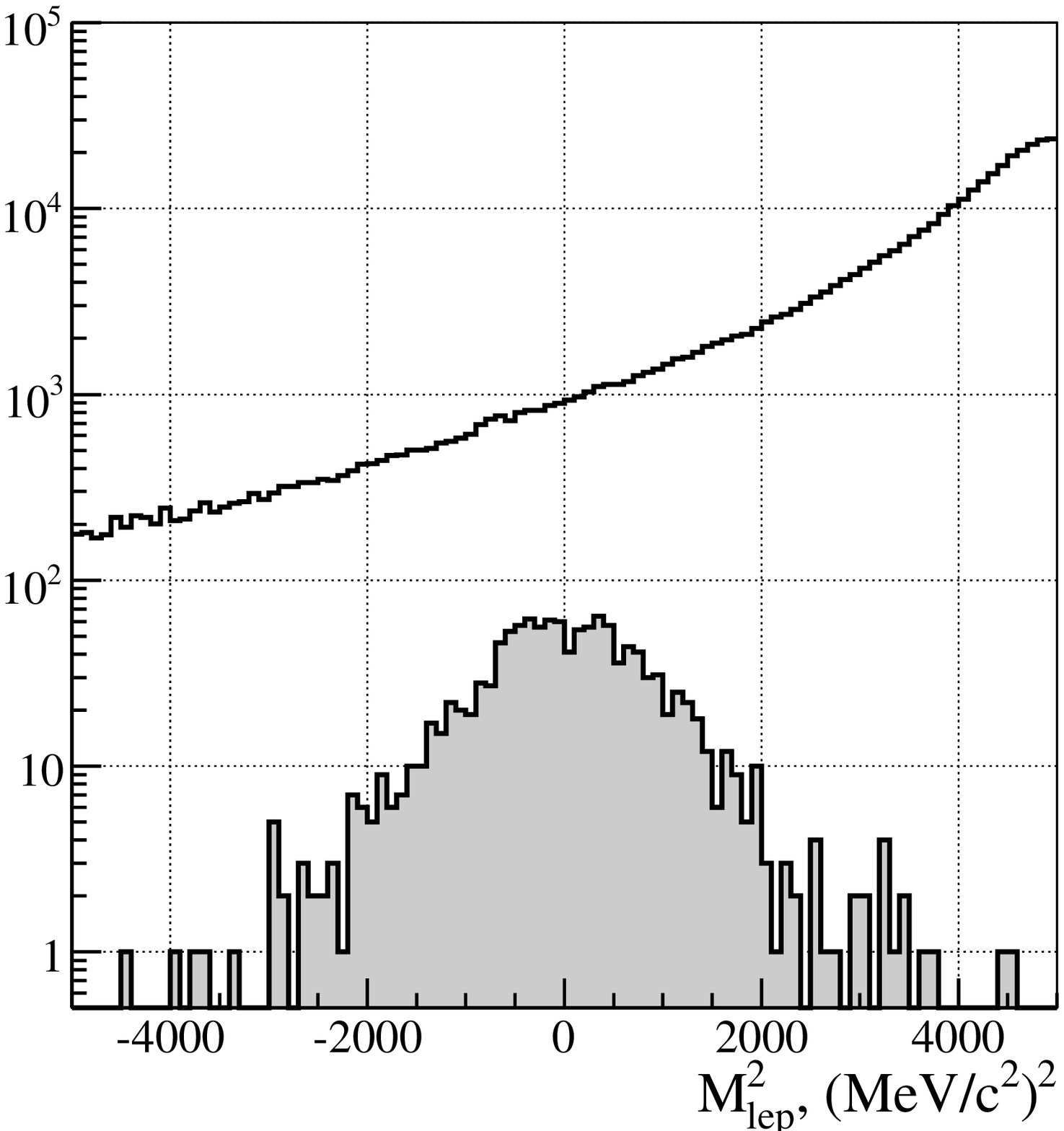}\\
\end{tabular}
    \caption{\label{ke2:mlep} MC distribution of $M_\mathrm{lep}^2$ before (left) and after
(right) quality cuts are applied. Shaded histogram: \ke events. Open histograms: background.
In MC, $R_K$ is set to the SM value.}
    \par\vspace{0pt}
  \end{minipage}\hspace*{0.2in}
  \begin{minipage}[c]{0.57\textwidth}
    \begin{tabular}[c]{cc}
  \includegraphics[totalheight=3.8cm]{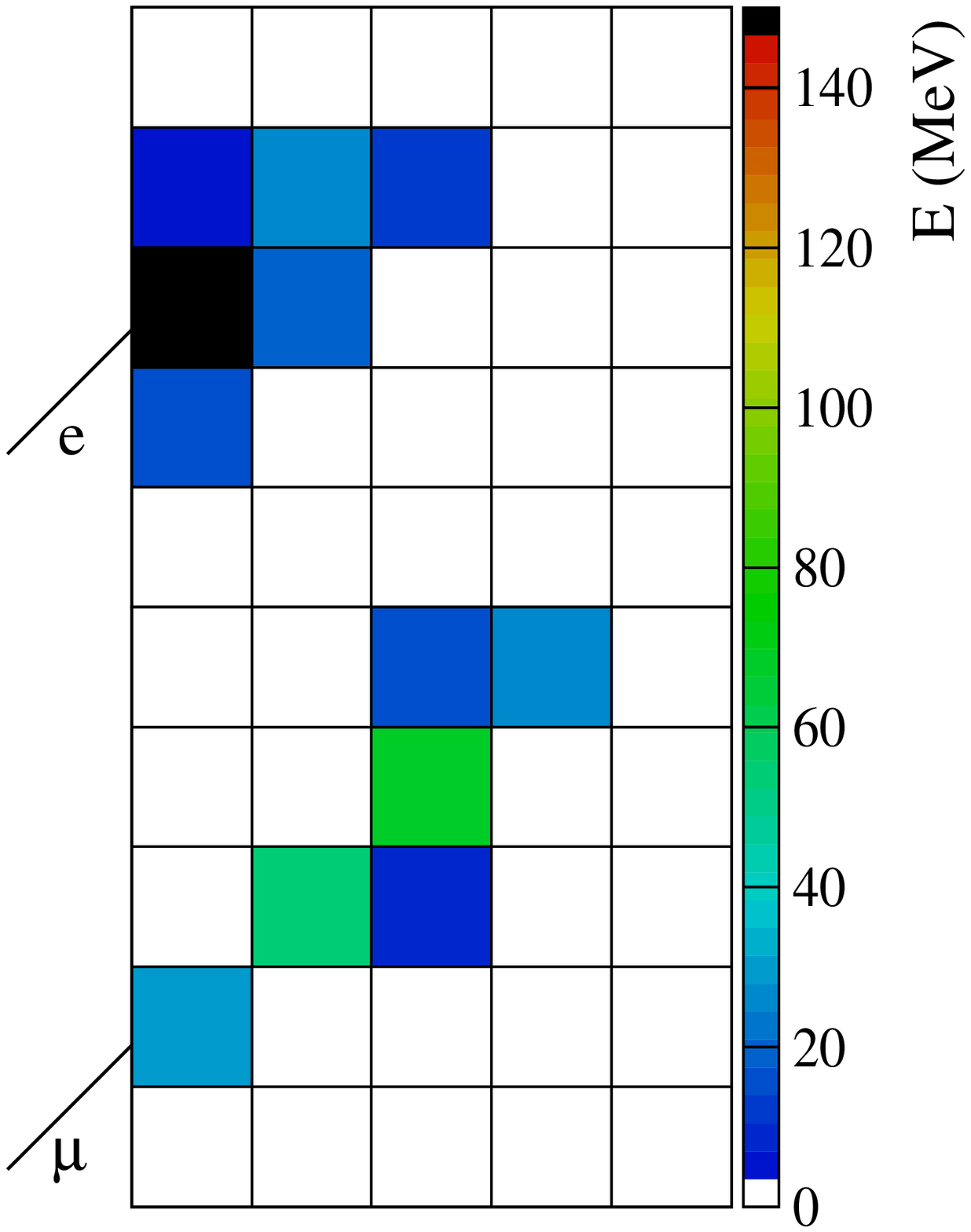}&
  \includegraphics[totalheight=3.6cm]{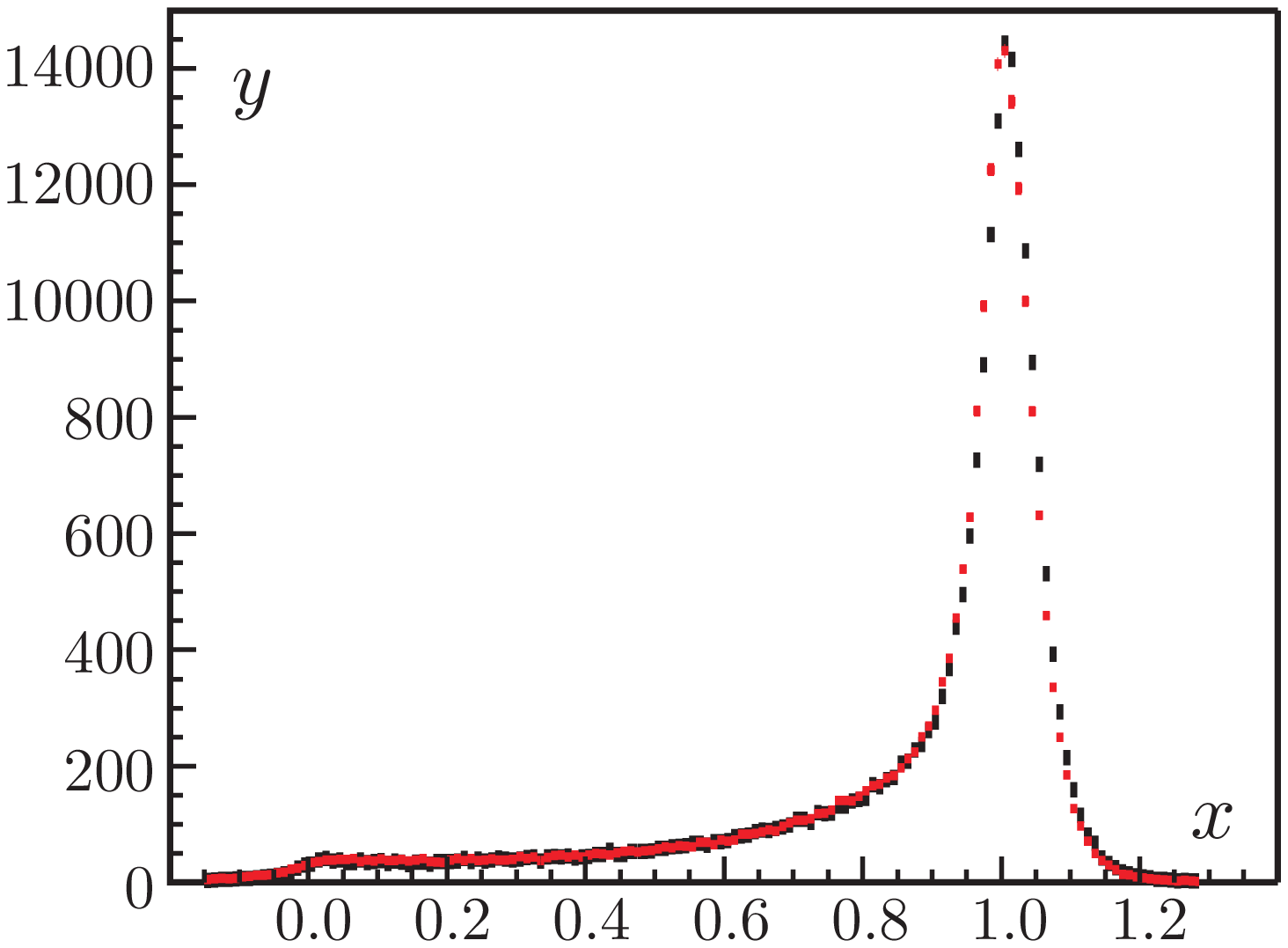}\\
\end{tabular}
    \caption{\label{ke2:display} Left: cell distribution for 200 MeV $e$ (top) and $\mu$
(bottom) from two selected events from $K_L\to \pi \ell \nu$.
Right: Distribution of NN output, \nn, for electrons of a $K_L\to
\pi e \nu$ sample from data (black histogram) and MC (red histogram).
}
    \par\vspace{0pt}
  \end{minipage}
%
\end{figure}

A powerful kinematic variable used to distinguish \ke and \km decays
from the background is calculated from the track momenta of the kaon and the
secondary particle: assuming $M_\nu=0$,
the squared mass of the secondary particle
($M_\mathrm{lep}^2$) is evaluated. The distribution of $M_\mathrm{lep}^2$ is shown
in Fig.~\ref{ke2:mlep} for MC events before and after quality cuts are
applied. The selection applied is enough for clean
identification of a \km sample, while further rejection is needed in order
to identify \ke events: the background, which is dominated by badly
reconstructed \km events,
is $\sim$10 times more frequent than the
signal in the region around $M_{e}^2$.

Information from the EMC is used to improve background rejection. To
this purpose, we extrapolate the secondary track to the EMC surface
and associate it to a nearby EMC cluster.  
For electrons, the associated cluster is close to the EMC surface
and the cluster energy $E_\mathrm{cl}$ is a
measurement of the particle momentum $p_\mathrm{ext}$, so that
$E_\mathrm{cl}/p_\mathrm{ext}$ peaks around 1.  
For muons, clusters tend to be more in depth in the EMC and $E_\mathrm{cl}/p_\mathrm{ext}$
tends to be smaller than 1, since only the kinetic energy is visible
in the EMC.  Electron clusters can also be distinguished from $\mu$
(or $\pi$) clusters, since
electrons shower and deposit their energy mainly in the first plane of
EMC, while muons behave like minimum ionizing particles in the first
plane and deposit a sizable fraction of their kinetic energy from the
third plane onward, when they are slowed down to rest (Bragg's peak),
see left panel of Fig.~\ref{ke2:display}.
Particle identification has been therefore based on the asymmetry
of energy deposits between the first and the next-to-first planes, on
the spread of energy deposits on each plane, on the
position of the plane with the maximum energy, and on
the asymmetry of energy deposits between the last and the next-to-last
planes. 
All information are combined with 
neural network (NN) trained on $K_L\to \pi \ell \nu$ data, taking into
account variations of the EMC response with momentum and impact angle
on the calorimeter.  The distribution of the NN output, \nn, for an
independent $K_L\to \pi e \nu$ sample is shown in the right panel of Fig.~\ref{ke2:display}
for data and Monte Carlo (MC). Additional separation has been obtained using time
of flight information.


The number of $\ke(\gamma)$ is determined with a binned likelihood fit
to the two-dimensional \nn vs $M_\mathrm{lep}^2$ distribution.
Distribution shapes for signal and \kmd background, other sources
being negligible, are taken from MC; the normalization factors for the
two components are the only fit parameters. 
In the fit region, a small fraction of $\ke(\gamma)$ events is due to
the direct-emission structure-dependent component (DE): 
the value of this contamination, $f_\mathrm{SD}$,
is fixed in the fit to the expectation from simulation.
This assumption has been evaluated by performing a dedicated
measurement of SD, which yielded as a by-product a determination of $f_{SD}$ with a 4\% accuracy.
This implies a systematic
error on $K_{e2}$ counts of 0.2\%, as obtained by repeating the fit with values of 
$f_\mathrm{SD}$ varied within its uncertainty.

\begin{figure*}
\begin{minipage}{0.55\linewidth}
  \begin{tabular}[c]{cc}
    \includegraphics[totalheight=3.8cm]{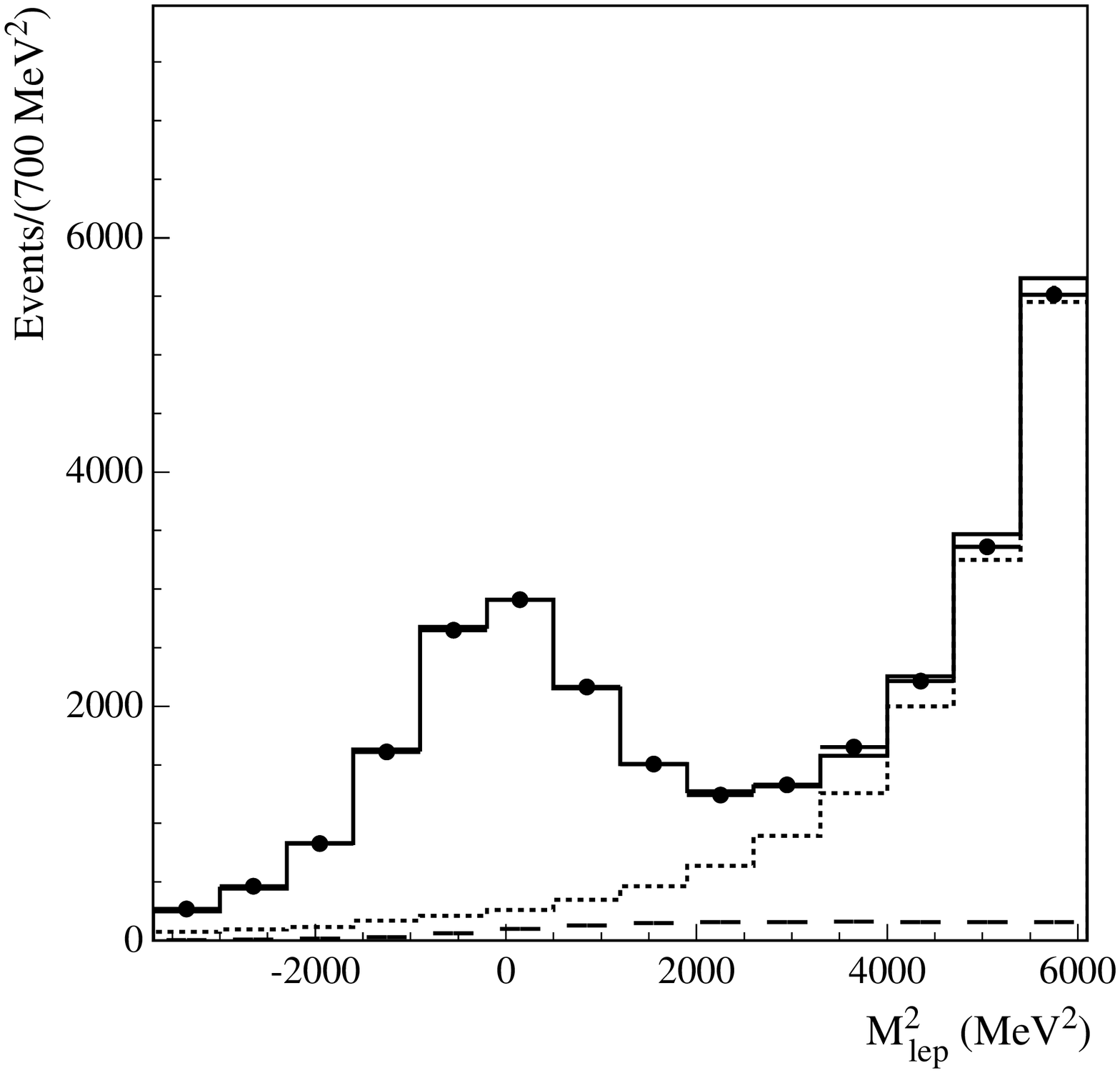} &
    \includegraphics[totalheight=3.8cm]{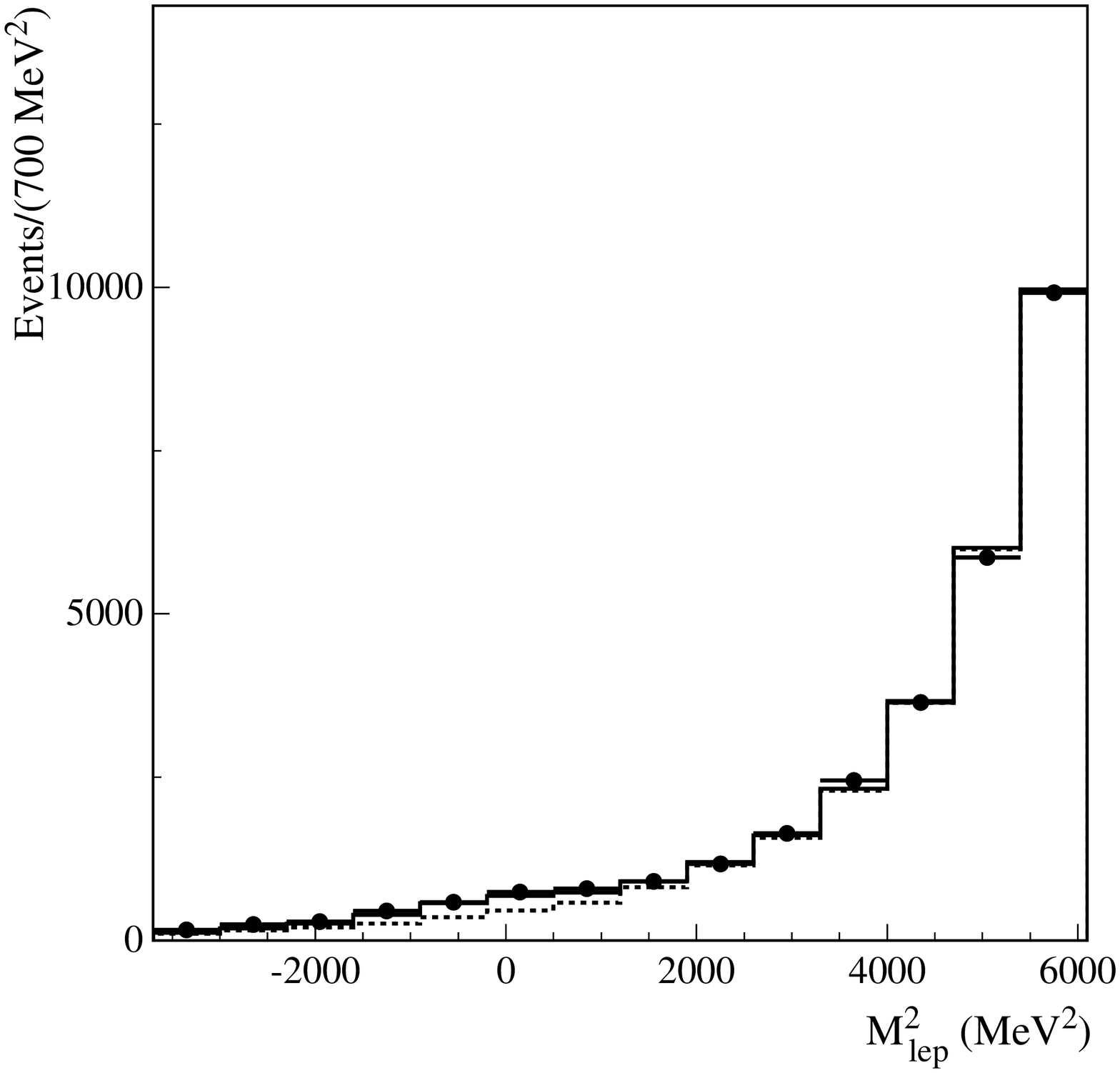} \\
  \end{tabular}
  \caption{\label{fig:fitke2}Fit projections onto the $M_\mathrm{lep}^2$ axis for two
    slices in NN output, $\nn>0.98$ and $\nn<0.98$, giving enhanced
    values of signal and background contributions, respectively.}
    \par\vspace{0pt}
  \end{minipage}\hspace*{0.2in}
  \begin{minipage}[c]{0.45\textwidth}
\centering
  \includegraphics[totalheight=3.8cm]{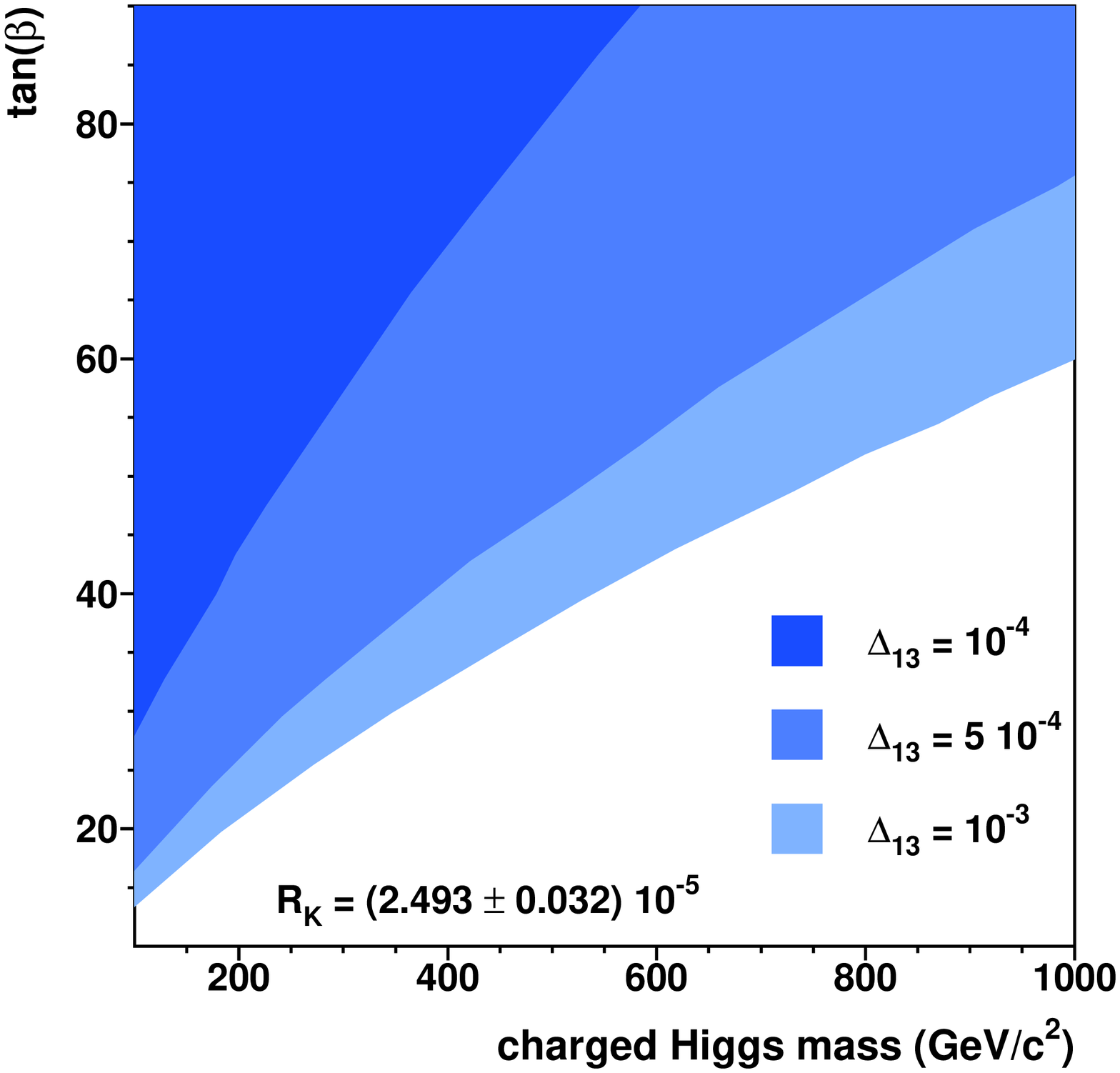}
  \caption{\label{fig:rkmssm} 95\%-CL excluded regions in the plane $\tan
    \beta$--charged Higgs mass for $\Delta_R^{31}=10^{-4},~5\times
    10^{-3},~10^{-3}$.}
\end{minipage}
\end{figure*}

In the fit region, we count 7064$\pm$102 $K^+\to e^+\nu(\gamma)$ and 6750$\pm$101 
$K^-\to e^-\bar{\nu}(\gamma)$ events.
Fig.~\ref{fig:fitke2} shows the sum of fit results for $K^+$ and
$K^-$ projected onto the $M_\mathrm{lep}^2$ axis in a signal-
($\nn>0.98$) and a background- ($\nn<0.98$) enhanced region. 

To assess the uncertainty on the $R_K$ measurement arising from
limited knowledge of the momentum resolution we have examined the
agreement between the $M_\mathrm{lep}^2$ distributions for data and MC
in the \kmd region.  For the \nn distribution, the EMC response at the
cell level has been tuned by comparing data and MC samples.
In order to evaluate the systematic error associated with these
procedures, we studied the result variation with different fit range
values, corresponding to a change for the overall $K_{e2}$ purity from
$\sim75\%$ to $\sim10\%$. The results are stable within
statistical fluctuations.  A systematic uncertainty of $0.3\%$ for
$R_{K}$ is derived ``\`a la PDG''~\cite{PDG+08} by scaling the
uncorrelated errors so that the reduced $\chi^2$ value of results is 1.

The number of \kmd\ events in the same data set is extracted from a
fit to the $M_\mathrm{lep}^2$ distribution. The fraction of
background events under the muon peak is estimated from MC to be $<0.1\%$.
We count $2.878\times10^8$ ($2.742\times10^8$)
$K^+_{\mu2}$ ($K^-_{\mu2}$) events.  
Difference in $K^+$ and
$K^-$ counting is ascribed to $K^-$ nuclear interactions in the material traversed.

The ratio of \ked\ to \kmd\ efficiency
is evaluated with MC and corrected for data-to-MC
ratios using control samples. To check the
corrections applied we also measured $R_3=\mathrm{BR}(\ket)/\mathrm{BR}(\kmt)$,
in the same data sample and
by using the same methods for the evaluation of the efficiency as for the $R_K$ analysis. 
We found $R_3=1.507(5)$ and $R_3=1.510(6)$, for $K^+$ and $K^-$ respectively. These are
in agreement within a remarkable accuracy with the expectation~\cite{Flavia+08}
from world-average form-factor slope measurements, $R_3=1.506(3)$.

\section{$R_K$ result and interpretation}

The final result is  $R_K=(2.493\pm0.025\pm0.019)\times 10^{-5}$. The 1.1\% fractional statistical 
error has contributions from signal count fluctuation (0.85\%) and background subtraction. 
The 0.8\% systematic error has a
relevant contribution (0.6\%) from the statistics
of the control samples used to evaluate corrections to the MC. 
The result does not depend on $K$ charge: quoting olny the uncorrelated errors,
$R_K(K^+)=2.496(37)10^{-5}$ and 
$R_K(K^-)=2.490(38)10^{-5}$.

The result in agreement with SM prediction of Eq.~\ref{eq:rkSM}. Including the new KLOE result, 
the world average reaches an accuracy at the \% level: $R_K=2.468(25)\times10^{-5}$. 
In the framework of MSSM with LFV
couplings, the $R_K$ value can be used to set constraints in the space of
relevant paremeters (see eq.~\ref{eq:rkmssm}).  The regions excluded
at 95\% C.L. in the plane $\tan \beta$--charged Higgs mass are shown
in Fig.~\ref{fig:rkmssm} for different values of the effective LFV
coupling $\Delta_R^{31}$.

\end{document}